\documentclass[epj,final]{svjour}

\usepackage{latexsym}
\usepackage{url}
\usepackage{amsfonts}
\usepackage{amsmath, amssymb}
\RequirePackage{graphicx}

\begin{document}
\title{The cosmological constant vs adiabatic invariance}
\author{Sh.Khlghatyan\inst{1}, A.A.Kocharyan\inst{2}, A.Stepanian\inst{1}, V.G.Gurzadyan\inst{1,3}
}                     
%
%
\institute{Center for Cosmology and Astrophysics, Alikhanian National Laboratory and Yerevan State University, Yerevan, Armenia \and 
School of Physics and Astronomy, Monash University, Clayton, Australia \and
SIA, Sapienza Universita di Roma, Rome, Italy}
\date{Received: date / Revised version: date}
%

\abstract{The property of adiabatic invariance is studied for the generalized potential satisfying the condition of identity of sphere's and point mass's gravity. That function contains a second term corresponding to the cosmological constant as weak-field General Relativity and enables to describe the dynamics of groups and clusters of galaxies and the Hubble tension as a result of two flows, local and global ones. Using the adiabatic invariance approach we derive the orbital parameters via Weierstrass functions, including the formula for the eccentricity which explicitly reveals the differences from the Kepler problem.} 
%
\maketitle

\section{Introduction}

The Hubble tension, i.e. the discrepancy in the values of the Hubble constant determined for the early and late Universe \cite{VTR,R1,R2,R3,Br} has stimulated the development of various approaches for its interpretation. The modified gravity models are among the actively investigated ones, also in the context of high precision tests of General Relativity (GR), see \cite{Cap2,E,C4,Cap3,Ciu,Dai} and references therein.

The weak-field modification of GR involving the cosmological constant is among the suggested approaches to the Hubble tension problem \cite{GS1,GS2}. It is based on a theorem proved in \cite{G1} that the general function for the force satisfying the identity of the sphere's and point mass's gravity field has the form
\begin{equation}\label{F}
\mathbf{F}(r) = \left(-\frac{\alpha}{r^2} + \Lambda r\right)\hat{\mathbf{r}}\ ,
\end{equation}   
where the second term in the r.h.s. corresponds to the cosmological term in McCrea-Milne cosmology \cite{MM,MM1}. In this form the cosmological constant $\Lambda$ enters the weak-field GR, and enables to describe the dynamics of groups and clusters of galaxies, as well as the flow of the Local Supercluster \cite{GS}. Namely, it appears that two Hubble flows can be distinguished, the local one described by the weak-field GR, and the global one described by Friedmannian equations \cite{GS1,GS2}. This is linked also with the problem of the formation of the cosmic web \cite{GFC} and the voids (e.g.\cite{spot}).

If Eq.(\ref{F}) reflects a fundamental physical law, then certain principal issues developed for the Newtonian potential can be reconsidered for that case as well. For example, it was shown that the appearance of the cosmological constant can be linked to a basic principle of nature, the time arrow \cite{AG}. The N-body problem for Eq.(\ref{F}) was studied in \cite{GKS} and was shown that the latter system becomes even more unstable with respect to that of Newtonian potential.   

Below we will inquire into a basic property of adiabatic invariance in the case of Eq.(\ref{F}), and corresponding modification of the properties of orbits in the two-body problem \cite{LL} in presence of the $\Lambda$-term and will derive the relevant orbital parameters.

\section{Action variables and Weierstrass functions}

The study of adiabatic invariance is an efficient way for revealing of the properties of Hamiltonian systems within the action-angle formalism \cite{LL,Arn,An}. Among the classical problems is the appearance of adiabatic invariants in the classification of orbits in the central field potential $V(r)=-\frac{\alpha}{r}$. In that case the Hamiltonian has the form  
\begin{equation}
H = \frac{1}{2}\left(p_{r}^2 + \frac{p_{\phi}^2}{r^2}\right)+ V(r).
\end{equation}
The action variables $I_{\phi}$ and $I_{r}$ are given as
\begin{equation}\label{KepJphi}
I_{\phi} = \frac{1}{2\pi}\int_0^{2\pi}p_{\phi}d\phi = L, 
\end{equation}
and
\begin{equation}\label{KepJr}
I_{r} = \frac{1}{2\pi}\oint p_rdr=-L + \frac{\alpha}{\sqrt{2|E|}}.
\end{equation}
The orbital parameters, the semi-latus rectum $l$ and the eccentricity $e$, can be expressed in terms of action variables as follows
\begin{equation}\label{Kepp}
l = \frac{I_{\phi}^2}{\alpha},
\end{equation}
\begin{equation}\label{Kepe}
e^2 = 1 - \left(\frac{I_{\phi}}{I_r + I_{\phi}}\right)^2.
\end{equation}  

Our aim is to derive the orbital parameters in the case of Eq.(\ref{F}), i.e. considering the contribution of the $\Lambda$ term. First, we need to find the action-angle  variables for the following Hamiltonian
\begin{equation}
H =\tfrac12g^{\mu\nu}p_{\mu}p_{\nu} =\tfrac{1}{2}\left(-\frac{p_{t}^2}{\Phi(r)}+\Phi(r)p_{r}^2+\frac{p_{\theta}^2}{r^2}+\frac{p_{\phi}^2}{r^2\sin^2\theta}\right),
\end{equation}
where $\Phi(r)=1-\frac{2M}{r}-\frac{\Lambda r^2}{3}$. Using the Hamilton's characteristic time-independent function \cite{LL}, we can write the action phase integral for $r$ as follows 
\begin{equation}
2\pi I_r=J_r=\oint_\gamma p_rdr=2\int_{r_1}^{r_2}\sqrt{E^2 -\Phi(r)\left(1+\frac{L^2}{r^2}\right)}\,\frac{dr}{\Phi(r)},
\end{equation}
where the motion $r_1\to r_2$ corresponds to positive $p_r$ and vice versa. Taking into account the smallness of the $\Lambda$, we expand $J_r$ in the series 
$$
J_{r}=J_{r}|_{\Lambda=0}+\left(\frac{\partial}{\partial\Lambda}J_r\right)\bigg|_{\Lambda=0} \Lambda+o(\Lambda).
$$ 
Then, using the substitution $z = \frac{2M}{r}-\frac{1}{3}$ and expanding the integrand into the sum of fractions, we obtain the Weierstrass form \cite{AIM,H} of the integral in the following form
\begin{align}\label{Jr}
J_r
&=2L\int_{e_2}^{e_1}\left[-1 
-\frac{A-B}{(z+\tfrac13)^2}
+\frac{B}{z+\tfrac13}
-\frac{B}{z-\tfrac23}\right]
\frac{dz}{\sqrt{P(z)}}\notag\\
\notag\\
&\qquad+L\,k^2\frac{\Lambda}{3}\int_{e_2}^{e_1}
\left[
-\frac{A - 2 B}{(z+\tfrac13)^4} 
-\frac{A - 4 B}{(z+\tfrac13)^3} 
- \frac{1 + A - 6 B}{(z+\tfrac13)^2}
- \frac{1 + A - 8 B}{(z+\tfrac13)}\right.\notag\\
&\qquad\qquad+\left.\frac{2B}{(z-\tfrac23)^2}
+\frac{1 + A - 8 B}{(z-\tfrac23)}\right]
\frac{dz}{\sqrt{P(z)}}+ o(\Lambda) \\
&:= J_{r}^0+\tfrac{4M^2\Lambda}{3} J_{r}^1 + o(\Lambda), \notag
\end{align}
where $A=\tfrac{4M^2}{L^2},\, B=E^2A$ and
\begin{align}
P(z)=z^3-\left(\tfrac13-A\right)z-\left(\tfrac{2}{27}+\tfrac{2}{3}A-B\right)=(z-e_1)(z-e_2)(z-e_3).
\end{align}
Since the discriminant is negative $\Delta=4g_2^3-27g_3^2<0$, then $e_1,e_2,e_3$ can be expressed in the following form for the three real roots
\begin{align}
e_m = 2\,\sqrt{\frac{g_2}{3}}\,\cos\left[\,\frac{1}{3} \arccos\left(\frac{3g_3}{2g_2}\sqrt{\frac{3}{g_2}}\,\right) - \frac{2\pi m}{3}\,\right] \qquad \text{for } m=1,2,3\,,
\end{align}
where $g_2=\frac13-A$ and $g_3=\tfrac{2}{27}+\tfrac{2}{3}A-B$.
To find each of the separate integrals
\begin{align}
\int_{e_2}^{e_1}\frac{1}{(z-\xi)^n}\frac{dz}{\sqrt{P(z)}}
\end{align}
in Eq.(\ref{Jr}), we use the  Weierstrass $\wp$-function \cite{H}. Making the substitution $z = \wp(w)$, we get
\begin{align}
F_n(x)&=\frac{1}{4\omega}\int_{e_2}^{e_1}\frac{1}{(z-\xi)^n}\frac{dz}{\sqrt{P(z)}}
=\frac{1}{4\omega}\int_{\omega+\omega'}^{\omega'}\frac{dw}{(\wp(w)-\wp(x))^n},
\end{align}
where $\xi = \wp(x)$ and $\omega,\omega'$ are the half periods of $\wp(w)$. Functions $F_n(x)$ are calculated using the following relations
\begin{align}
F_1(x)&=-\frac{Q(x )}{\wp'(x )};\\
F_2(x)&=\frac{Q(x ) \wp''(x )+S(x ) \wp'(x )}{\wp'^{3}(x )};\\
F_3(x)&=\frac{-3 Q(x) \wp''^{2}(x )+\wp'^{3}(x )+\wp'(x ) \wp'''(x ) Q(x )-3 S(x ) \wp''(x )}{2\wp'^{5}(x)};\\
F_4(x)&=\frac{1}{6\wp'^{7}(x)}\left[15 Q(x) \wp''^{3}(x) 
-5 \wp'^3(x) \wp''(x) -\wp'^2(x) \left(4 \wp^{(3)}(x) S(x)-\wp^{(4)}(x) Q(x)\right)\right.\notag\\ 
&\qquad\left.+5 \wp'(x) \wp''(x) \left(3 S(x) \wp''(x) +2 \wp^{(3)}(x) Q(x)\right)\right],
\end{align}
where
\begin{align}
Q(x)&=\zeta(x)-\frac{\eta}{\omega} x,\\
S(x)&=-\frac{\eta}{\omega} -\wp(x).
\end{align}
Here $\zeta(x)$ is the Weierstrass $\zeta$ function and $\eta$ is its quasi-period. Using these expressions and taking into account that by definition $\wp^{'2}(z)=4P(\wp(z))$ from which it follows, in particular, that
\begin{align}
A=\frac{1}{2}\wp''(a)\,,\quad    B=\frac{1}{4}\wp'^2(b),
\end{align}
where $\wp(a)=-\tfrac13$ and $\wp(b)=\tfrac23$, we can write the integrals  $J_r^0, J_r^1$ in the following form 
\begin{align}
J_r^0&=
4\omega L\left[-1  +\wp'(b)Q(b)
+\frac{1}{\wp'(a)}\left(\frac{\wp''^2(a)}{\wp'^2(a)} -\wp'^2(b)\left(1-\frac{\wp''(a)}{\wp'^2(a)}\right)\right)Q(a)
-\left(\frac{\wp'^2(b)}{\wp'^2(a)}+\frac{\wp''(a)}{\wp'^2(a)}\right)S(a)
\right];\\
J_r^1 &= 2\omega L\Bigg[2 S(b)+\frac{4 \wp'^2(b)-\wp''(a)}{2 \wp'^2(a)}+\frac{5 \wp''(a) \left(-2 \wp'^2(b)+\wp''(a)\right)}{6 \wp'^6(a)}+\frac{Q(b) \left(-1+8 \wp'^2(b)-\wp''(a)+2\wp''(b)\right)}{\wp'(b)}\nonumber\\
&\quad +\frac{1}{6 \wp'^6(a)}S(a) \Bigg(6 \wp'^4(a) \left(-1+6 \wp'^2(b)-\wp''(a)\right)+9 \wp'(a) \wp''(a) \left(-4 \wp'^2(b)+\wp''(a)\right)
+\left(2\wp'^2(b)-\wp''(a)\right)\nonumber\\
&\quad \left(15 \wp''^2(a)-4 \wp^{(3)}(a)\right)\Bigg)
+\frac{1}{6 \wp'^7(a)}Q(a) \Bigg(15 \left(2 \wp'^2(b)-\wp''(a)\right) \wp''^3(a)+6 \wp'^6(a)\left(1-8 \wp'^2(b)+\wp''(a)\right)\nonumber\\
&\quad -6 \wp'^4(a) \wp''(a) \left(1-6 \wp'^2(b)+\wp''(a)\right)+9 \wp'^2(a) \wp''^2(a) \left(-4 \wp'^2(b)+\wp''(a)\right)
\nonumber\\
&+3 \wp'^3(a) \left(4\wp'^2(b)-\wp''(a)\right) \wp^{(3)}(a)
 +\wp'(a) \left(2 \wp'^2(b)-\wp''(a)\right) \left(10 \wp''(a) \wp^{(3)}(a)+\wp^{(4)}(a)\right)\Bigg)\Bigg].
\end{align}
These expressions can be simplified if we use the following series \cite{H}
\begin{align}
	\wp(x)&=-\frac{\eta}{\omega}+\left(\frac{\pi}{2\omega}\right)^2\sin^{-2}\frac{\pi x}{2\omega}
	-8\left(\frac{\pi}{2\omega}\right)^2\sum\limits_{n=1}^\infty\frac{nq^{2n}}{1-q^{2n}}\cos\frac{n\pi x}{\omega}\\
	\frac{\eta}{\omega}& = \frac{1}{3}\left(\frac{\pi}{2\omega}\right)^2 -8\left(\frac{\pi}{2\omega}\right)^2\sum\limits_{n=0}^\infty\frac{q^{2(2n+1)}}{(1+ q^{2(2n+1)})^2},\\
	\sqrt{\frac{2\omega}{\pi}}
	&=\frac{2}{(e_1-e_3)^{1/4}+(e_1-e_2)^{1/4}}(1+2q^4+2q^{16}+\cdots),\\
	Q(x)&
	=\frac{\pi}{2\omega}\cot\frac{\pi x}{2\omega}
	+\frac{2\pi}{\omega}\sum\limits_{n=1}^\infty\frac{q^{2n}}{1-q^{2n}}\sin\frac{n\pi x}{\omega},\\
	S(x)& =-\left(\frac{\pi}{2\omega}\right)^2\sin^{-2}\frac{\pi x}{2\omega}
	+8\left(\frac{\pi}{2\omega}\right)^2\sum\limits_{n=1}^\infty\frac{nq^{2n}}{1-q^{2n}}\cos\frac{n\pi x}{\omega},
\end{align}
where $q=e^{i\frac{\pi\omega'}{\omega}}$.

\section{Orbital parameters}

Based on the above derivations, we are now able to express, as in the Keplerian case, Eqs.(\ref{Kepp}-\ref{Kepe}), also for the case of Eq.(\ref{F}), the orbital parameters using the action variables
\begin{align}
l&=\frac{I_\phi^2}{M} -\left(e^2+3\right) M\,,\\
\frac{I_r}{I_\phi}
&=-1+\frac{1}{I_\phi^2-\left(e^2+3\right) M^2}\left(\frac{I_\phi^2}{(1-e^2)^{1/2}}-3 M^2\right)
+\frac{I_\phi^6}{\left(1-e^2\right)^{7/2} M^2} \left(\frac{3 e^2+2}{2 M^2}+\frac{7 e^4+22 e^2+6}{\left(e^2+3\right) M^2-I_\phi^2}\right)\frac{\Lambda }{3}\,.
\end{align}
For the eccentricity one gets the first order to $\Lambda$ approximation in the following form
\begin{align}
e^2&\approx1-\nu ^2+\frac{M^2}{I_\phi^2} \left(\nu ^2+3 \nu -4\right) \nu ^2
+\frac{I_\phi^4}{M^4\nu ^4}\left[I_\phi^2(3 \nu ^2-5)+M^2 \left(20 \nu ^4+27 \nu ^3-140 \nu ^2-75 \nu +190\right)\right]
\frac{\Lambda}{6},
\end{align}
where
\begin{align}
\nu=\frac{I_{\phi}}{I_r + I_{\phi}}.
\end{align}
This reveals an important difference from the Keplerian case (cf. with the post-Newtonian corrections in \cite{Ch}). If in the classical Kepler problem the eccentricity remains unchanged at a slow change of $M$ due to adiabatic invariance of $I_{\phi}$ and $I_r$, in this case the eccentricity becomes dependent on $M$.

Figure 1 exhibits the behavior in time of the action $I_r$ and of the eccentricities of orbits for Newtonian $e(t)$ and $\Lambda$-potentials $e_{\Lambda}(t)$. The variation of the difference of eccentricities $\Delta e= e(t) - e_{\Lambda}(t)$ (Fig.2) indicates the increase of the role of $\Lambda$-term for the large scale finite orbits.

\begin{figure}[h!]
	\caption{The variation of action $I(r)$ and of eccentricities for Newtonian $e(t)$ and $\Lambda$-potentials $e_{\Lambda}(t)$ for the given input parameters.}
	\centering
	\vspace{5mm}
	\includegraphics[scale=0.7]{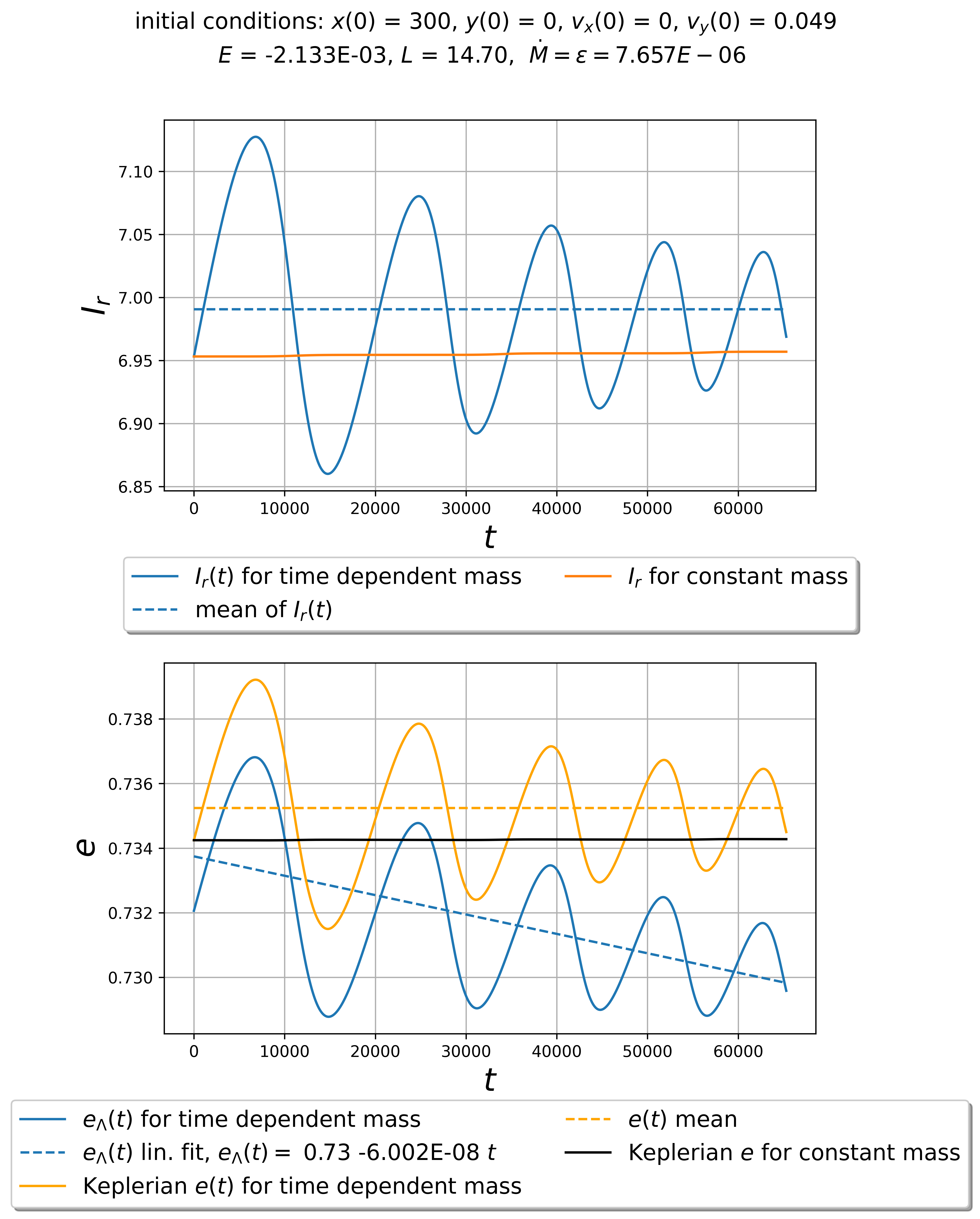}
\end{figure}
\begin{figure}[h!]
	\caption{The variation of the difference of eccentricities $\Delta e = e(t) - e_{\Lambda}(t)$ for 3 scales of the orbits.}
	\centering
	\vspace{5mm}
	\includegraphics[scale=0.7]{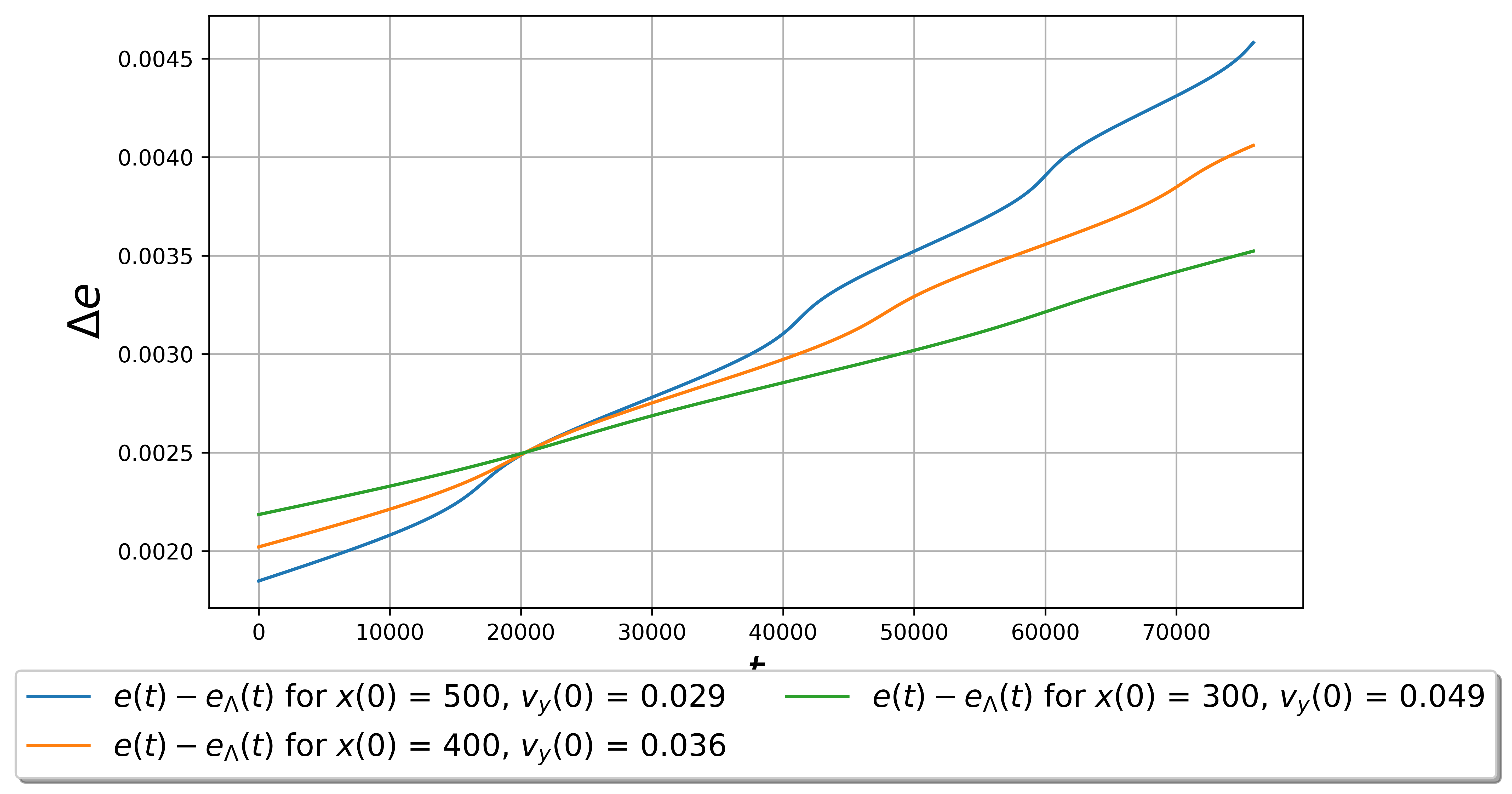}
\end{figure}

\section{Conclusions}

The adiabatic processes and adiabatic invariance are known as important concepts in classical mechanics, quantum mechanics, thermodynamics \cite{LL,Arn,An}. We used that approach to study the properties of orbits for the gravity field given by Eq.(\ref{F}), as compared with the Kepler problem for the Newtonian potential. Based on the fact that the $\Lambda$ constant is of a small value \cite{G1,GS}, we obtain the first order formulae (to $\Lambda$) of the corresponding expansion series by means of Weierstrass functions. In the same approximation, i.e. for first order to $\Lambda$, we derive the formula for the eccentricity via the actions which reveals the genuine difference of those orbits from the Keplerian case.    

Previously, it was shown the remarkable role of the cosmological constant term in Eq.(\ref{F}) in the irreversibility and dynamic entropy
production on the large scales \cite{AG}. Considered as a vacuum energy \cite{Z}, the cosmological constant (dark energy) can also serve as an ideal thermodynamic bath without any back-reaction. Regarding the exponential instability (chaos) of N-body gravitating systems, the systems with interaction of Eq.(\ref{F}) are shown to be even more unstable than those with Newtonian potential \cite{GKS}. The consideration of the adiabatic invariance and the derivation of the orbital parameters presented above reveal further principal properties for the systems with the $\Lambda$-term interaction.

\end{document}